# Optimizing the Optical and Electrical Properties of Graphene Ink Thin Films by Laser-annealing

*Sepideh Khandan Del[1], Rainer Bornemann[2], Andreas Bablich[1], Heiko Schäfer-Eberwein[2], Jiantong Li[3], Torsten Kowald[4], Mikael Östling[3], Peter Haring-Bolívar[2], Max. C. Lemme[1*]*

[1]University of Siegen, Institute of Graphene-based Nanotechnology, Hölderlinstr. 3, D-57076, Siegen, Germany

[2]University of Siegen, Institute of High Frequency and Quantum Electronics, Hölderlinstr. 3, D-57076, Siegen, Germany

[3]KTH-Royal Institute of Technology, School of Information and Communication Technology, Electrum 229, SE-164 40 Kista, Sweden

[4]University of Siegen, Institut of Materials Chemistry, Paul-Bonatz-Str. 9-11, D-57076, Siegen, Germany

[*]corresponding author: max.lemme@uni-siegen.de

**Abstract**

We demonstrate a facile fabrication technique for graphene-based transparent conductive films. Highly flat and uniform graphene films are obtained through the incorporation of an efficient laser annealing technique with one-time drop casting of high-concentration graphene ink. The resulting thin films are uniform and exhibit a transparency of more than 85% at 550 nm and a sheet resistance of about 30 kΩ/□. These values constitute an increase of 45% in transparency, a reduction of surface roughness by a factor of four and a decrease of 70% in sheet resistance compared to unannealed films.

**Main Text**

Graphene dispersions are potential candidates for a variety of applications in printed electronics and as surface coatings. Transparent conductors made from graphene dispersions show promising properties compared to traditionally used metallic oxide thin films. In contrast to many metal oxides, graphene-based transparent films are flexible, chemically stable and exhibit uniform broadband absorption/transmission. In addition, their production may ultimately be more cost efficient. Recently, many groups investigated transparent graphene conductors made from graphene dispersions by inkjet printing,[1–3] vacuum filtration,[4,5,6] , , and spray coating.[7] Here, we propose a laser annealing technique for the fabrication of high performance transparent conductive films made from graphene dispersion. This technique is simple, fast and on demand in comparison with other methods commonly used. In addition, this technique can be applied for lithographic definition of transparent thin film conductors made from graphene ink. Uniform, transparent and conductive graphene thin films were fabricated by simple drop casting of graphene dispersion combined with a laser annealing process. The laser annealing method is primarily investigated with drop-casted graphene dispersions and its universality is confirmed with inkjet printed graphene films.

Graphene ink with multiple graphene layers was produced by a solvent exchange technique,[8] where graphite powders were first exfoliated in dimethylformamide (DMF) and stabilized with Ethyl Cellulose. Then, DMF was exchanged by terpineol to increase the graphene concentration, as well as to adjust the ink viscosity and to reduce solvent toxicity.[9] The graphene concentration in the final inks is as high as around 1 mg/mL.

In order to form multilayer graphene films, the graphene ink was dispersed on a glass slide by controlled drop-casting (1µL droplets). The high-concentration inks allow efficient fabrication of graphene films through one-time drop casting. After ink drying, the obtained films were baked at 400 °C for 30 minutes to remove the stabilizing polymer which is necessary to prevent graphene flake aggregation during ink drying. Subsequently, the films were treated by scanning with a continuous wave laser beam (500 mW, 532 nm) across the surface with a power density of 55 MW/cm$^2$ for a duration of 1 to 4 ms and a spot size of 1 μm. **Figure 1a** shows a schematic of the laser treatment process. In addition, this laser system was set up to conduct Raman measurements to enable in situ characterization throughout the process, albeit at lower laser power.

The graphene thin films were characterized by a combination of optical microscopy, laser scanning microscope, scanning electron microscope (SEM), Raman spectroscopy and X-ray powder diffraction (XRD). Moreover, the electrical properties and the transparency of the films were characterized.

**Figure 1b** shows an optical micrograph of the film, including the areas where the laser annealing process was performed (marked by the dashed rectangles). In contrast to the untreated areas, the optical image shows improved transparency and homogeneity after laser annealing.

The topography of the films was studied with a KEYENCE VK-X 200 laser scanning microscope (**Figure 1c**). The roughness of the laser treated film ($R_{RMS}$) is about 11 nm, which is about four times smaller than that of as-deposited film (~ 60 nm). The average step height between the two areas in Figure 1c is approximately 200 nm. These two results indicate a severe flattening effect of the laser annealing process on the multilayer graphene flake film.

We carried out a systematic study of the influence of the laser annealing power. Different areas of drop-casted graphene films were exposed to the laser with different power and exposure times. The results are shown in **Figure 2a**, where the power ranges from 55 MW/cm$^2$ to 17 MW/cm$^2$ and the exposure time ranges from 4 ms to 1 ms. At 17 MW/cm$^2$, there is no optically discernible effect of the laser on the film. We therefore chose to carry out Raman spectroscopy for both annealed and unannealed films at this laser power. **Figure 2b** shows the average transmittance for light with a wavelength of $\lambda = 550$ nm for the unannealed (light blue background) and the annealed regions of the film (patterned logo of the University of Siegen). The transparency of the film increased from 40% to over 85% after the laser annealing process.

SEM images (**Figure 3a**) of annealed and unannealed films were taken under a tilt angle of 52° to enhance surface topography. The insets show the respective regions with higher magnification. These images demonstrate the topographical changes in the film due to the laser annealing process. In the annealed region, graphene flakes are finer, smoother and more compact, which results in a significantly improved surface homogeneity.

**Figure 3b** shows Raman spectra of the graphene ink drop-casted on a glass slide before and after laser annealing, but including thermal annealing at 400°C in both cases. In order to obtain Raman data for the unannealed area, a low laser power was chosen, i.e. less than the threshold power required for a visible annealing effect as discussed in Figure 2. The spectra clearly show the signatures of graphitic films, i.e. the G and the 2D bands. In addition, we observe D and D' intensities in the spectra. While the G peak is linked to the relative motion of sp$^2$ carbon atoms and indicates the presence of graphene, the D peak is due to the breathing modes and requires defects for its activation (double resonance effect).[10–12] The 2D peak is the second order of the D peak and is always seen even in the absence of the D peak.[11] The origin of the D' peak is due to an intravalley double resonance effect.[13] Both the D and the D' peak can be attributed to either the presence of a considerable amount of defects within the graphitic flakes or to disorder at the edges of flakes.[13,14] The full width at half-maximum of the G peak, FWHM (G), increases with structural disorder.[15,16] Moreover, the intensity ratio I(D)/I(G) increases with increasing edge disorder.[15] Therefore, combining I(D)/I(G) with FWHM(G) can help to discriminate between disorder localized at the edges and disorder within the flakes. In the latter case, changes in I(D)/I(G) would correlate with FWHM(G). In the laser annealed samples, both FWHM(G) and I(D)/I(G) decreased, from 40 cm$^{-1}$ to 27 cm$^{-1}$ and from 0.62 to 0.34 respectively, indicating that the laser annealing process significantly reduces structural disorder in the film.

XRD spectra were recorded to investigate the effect of the laser annealing on the structure of the film. The results are shown in **Figure 3c.** The peak at 2Θ = 26.7° corresponds to the 002 plane of graphite layers with an interlayer spacing of 0.334 nm. The intensity of this peak is higher for laser treated samples, indicative of a higher degree of graphene flake overlap in the (002) crystalline direction. This higher XRD count supports the hypothesis of a flattening effect of the laser annealing process that was posed based on the SEM images and the laser scanning microscope data.

The sheet resistance of the graphene films was assessed with a 4-point probe technique.[17-18] The un-annealed films exhibit a high sheet resistance of approximately 1MΩ/□. The laser treatment process reduces this value considerably by three to four orders of magnitude to approximately 30 kΩ/□. We have established by analytical methods that the laser annealing process results in a flattening effect of the graphene layers. The considerable reduction of sheet resistance confirms this assessment: electrical transport in graphene films is dominated by the inter-flake junction resistance and the number of junctions in the percolation paths.[19] As the graphene films are flattened, the flake overlap / the junction area increases (compare SEM of untreated film in Figure 3a) and the inter-flake resistance decreases drastically. Moreover, since most graphene lies flat on the substrate, much less flakes are needed to form a percolation path. As a result, the film sheet resistance is significantly decreased.

In addition to flattening the graphene films, laser annealing may play a role in removing remaining polymer (ethyl cellulose) trapped between the flakes by local heating. The heat flow in the cross-plane direction of the graphene layers and graphite flakes is strongly limited by weak inter-plane van der Waals forces. The drop-casted films mainly consist of randomly oriented graphene flakes. Therefore, in line with the poor electrical conductivity, we conclude that the heat induced by the laser hardly dissipates, giving rise to a rapid local temperature increase in the film.

The removal of polymer likely contributes to the dramatic thickness decrease of the film after annealing: there is a step of 200 nm between the laser annealed area and the un-annealed area in Figure 1c. Such a loss of material would certainly contribute to improved transparency, in addition to an improvement due to better flake alignment and – consequently - reduced scattering. Polymer removal would also contribute to a reduced sheet resistance in the flake percolation network.

Finally, the laser may burn highly defective parts of the graphene flakes, since the experiments have been carried out in ambient air. This would result in an overall improved sheet resistance. Please note that while there is no direct experimental evidence for this aspect, a chemical vapor deposited graphene film (with much less defects) was annealed by the laser in a control experiment. This film did not change its conductive properties after annealing.

Figure 4a compares the sheet resistance and transmittance of the drop-casted graphene films before and after laser annealing. The laser treatment improves both electrical and optical properties of the film significantly. In order to verify the universality of the laser annealing technique, we performed an experiment on inkjet printed graphene films (6 printing passes) based on the same ink.[2] The result is similar, if less pronounced, with a decrease in sheet resistance from ~180 kΩ/□ to ~60 kΩ/□ and an increase in transmittance from 52% to 54% (λ = 550 nm).

The inkjet printed films are initially smoother and more homogenous than drop casted samples. AFM measurements show an initial surface roughness for the as printed sample of about 37 nm (RMS), and this value drops to about 26 nm after laser annealing. In contrast, the roughness (RMS) of drop casted samples dropped from 53 nm to 22 nm through laser annealing. This serves to explain the lower degree of improvement in the inkjet printed films, as these have less potential for further improvement of the optical and electrical properties.

Table 1 compares the sheet resistance and the transparency of the reported graphene films at λ = 550 nm with literature data fabricated from graphene dispersion with various methods. These include inkjet printing of graphene ink with further thermal annealing at 400°C [2–3], vacuum filtration through a polytetrafluoroethylene membrane [4] vacuum filtration through mixed cellulose ester membranes (Millipore) and further annealing at 250 °C [5], vacuum filtration with an alumina membrane followed by thermal annealing at 250°C in Argon atmosphere [6] and spray coating with subsequent thermal annealing at 250°C in Argon atmosphere [7] The films fabricated by the method in this work exhibit electrical and optical properties comparable to that of the state of the art.

In conclusion, we investigated transparent conductive graphene thin films fabricated by drop-casting of graphene ink, baking and subsequent annealing with a scanning laser beam. The laser annealed films show low sheet resistance and high transmittance, which make them suitable candidates for electronic applications such as transparent electrostatic dissipation.[20] The laser annealing technique is transferrable to graphene inks obtained with other methods, as demonstrated with inkjet printed graphene films. This work introduces a fast, on-demand and economical method to improve graphene ink properties and also to controllably transform an insulating coating into a transparent conductive film. Moreover, it provides the opportunity for large-scale fabrication and, if a pattern generator is used, a basis for lithographically defined transparent conductors. Our work may lead to the development of a new generation of transparent conductive films with thermal and chemical stability and low production cost.

**Methods**

The graphene inks were prepared through a solvent exchange technique as detailed in Li et al.[9] a mixture of 2 mg/mL of natural graphite flakes (Sigma–Aldrich, Product No. 332461) and dimethylformamide (DMF) was sonicated in a Branson 2510E-MTH bath ultrasonicator for 20 h, followed by centrifugation for 15 min at >10,000 rpm to sediment thick flakes. The supernatant was harvested and a small amount of polymer (ethyl cellulose) was added to protect the graphene flakes from agglomeration. The solution was then mixed with terpineol. A vacuum distillation process to ensure that distillation occurs at a temperature lower than the solvents' boiling points was used to remove DMF: The dispersion was heated to 80°C in a water bath. As the pressure was reduced to 30 mbar, only the DMF evaporated. After the DMF was boiled off, the remaining graphene/terpineol dispersion was harvested. Finally, a rough sonication for several minutes sufficed to obtain pure, particle-free graphene dispersion.

The Raman spectra and the transmission image were taken with a self-built Raman micro spectrometer, which can also be used for multispectral image acquisition. The optical setup is built around a commercial inverted microscope (TE 300, Fa. Nikon). For Raman excitation and the laser treatment, a green CW DPSS laser (532 nm, 500 mW, Fa. CNI) was used. The laser was coupled in the optical setup in the path for epi-fluorescence illumination. For multispectral transmission imaging the original tungsten lamp was used as a broad band light source. The spatially resolved spectra were taken with an imaging spectrograph and a liquid nitrogen cooled CCD camera (Triax 320 with a synergy CCD camera, Fa. Horiba JobinYvon). This detection part was attached to the microscope at the back reflection port. For the annealing and the imaging the samples were sampled stepwise with a motorized scanning stage (SCAN IM 120x100, Fa. Märzhäuser). The stepper motors of this microscope stage are trigged by the CCD camera, which acts as the master pixel clock for the complete image acquisition.

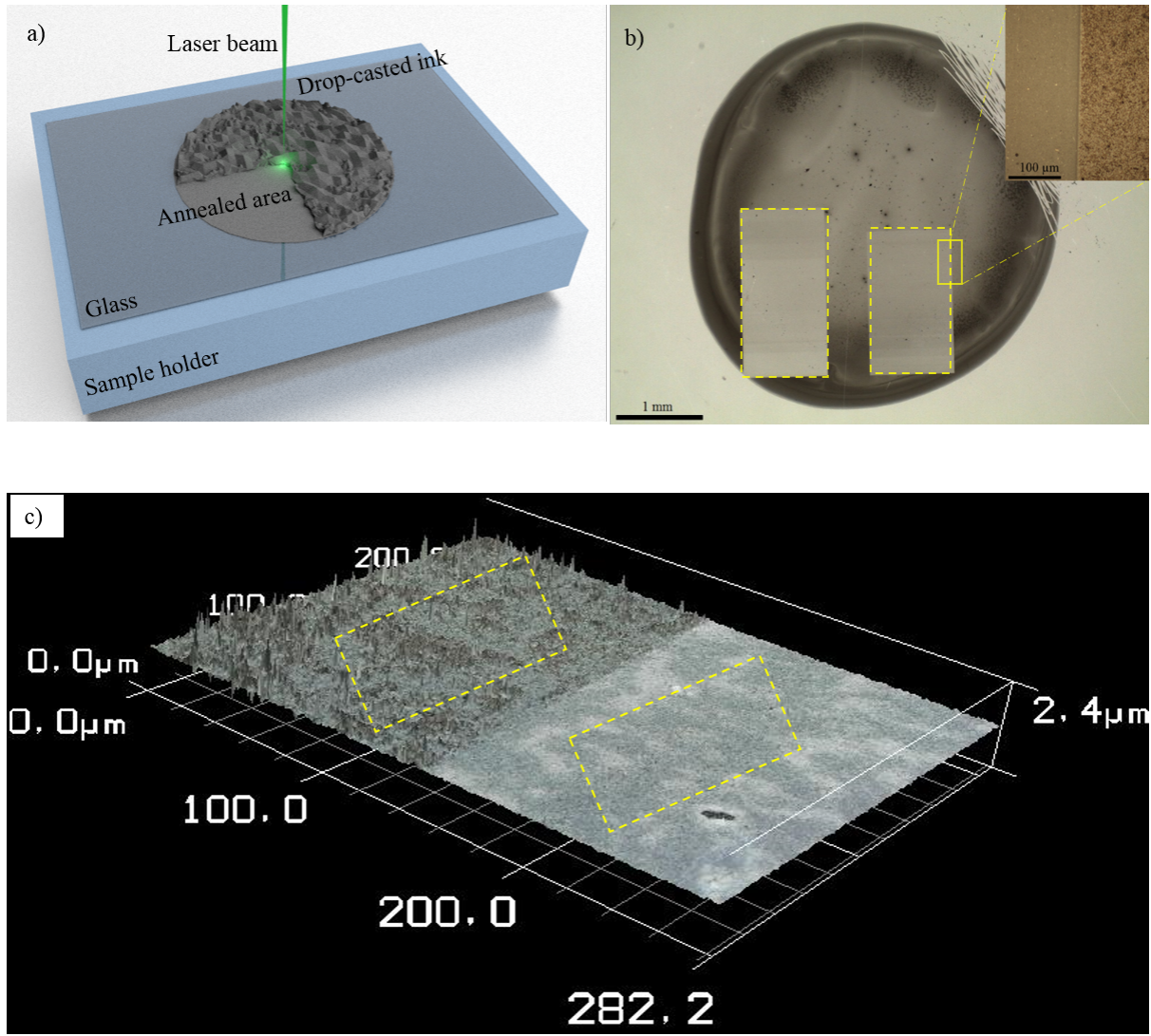

Figure 1. (a) Schematic of the laser annealing process, where a green laser scans across the drop-casted multilayer graphene film. (b) Optical microscope image of as-deposited film, laser treated areas are marked by dotted rectangle. The inset shows the border area with higher magnification (c) 3D image of the surface topography by Laser scanning microscope, the step size between 2 dotted rectangles is around 200nm which indicates the difference between average heights in the two areas.

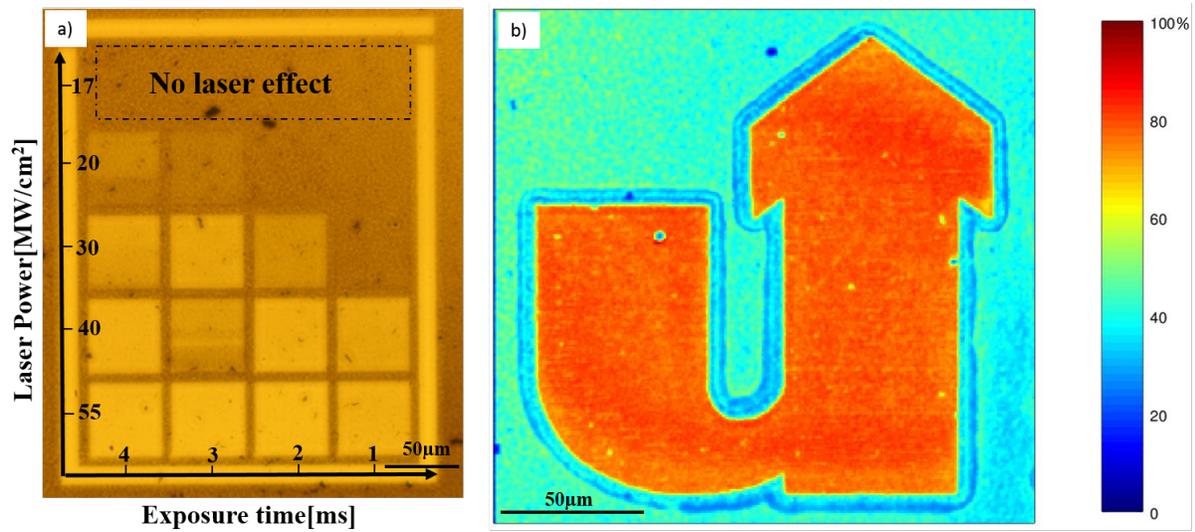

Figure 2. (a) Optical micrograph of a drop-casted graphene film exposed to a laser (λ = 532 nm) with different power and exposure time. (b) Laser lithography: Transparency map of the graphene film. The patterned University of Siegen logo (red) represents the laser annealed film while the background (green/blue) corresponds to the un-treated film.

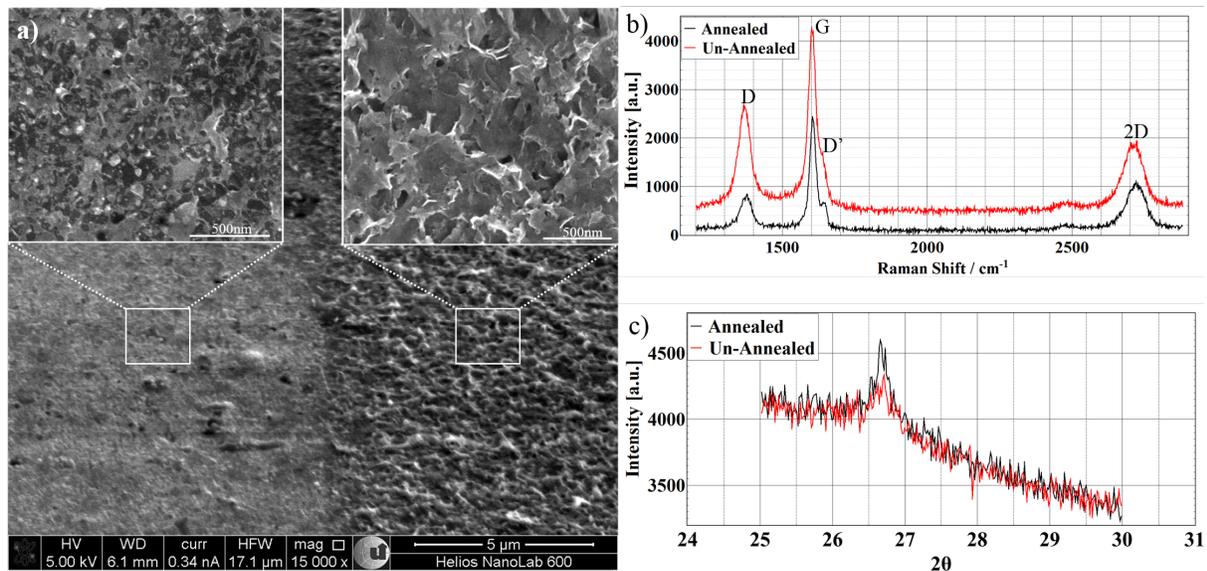

Figure 3. (a) SEM image of treated and un-treated areas, the insets show the areas with higher magnification on the order of 150000 with horizontal field width on the order of 2µm. (b) Raman Spectra of the un-treated (red) and the laser annealed film (black). The full width at half-maximum of the G peak and the intensity ratio of D and G peak decrease after laser annealing, indicating reduced structural disorder after annealing. (c) XRD diffractogram of laser annealed and un-treated films.

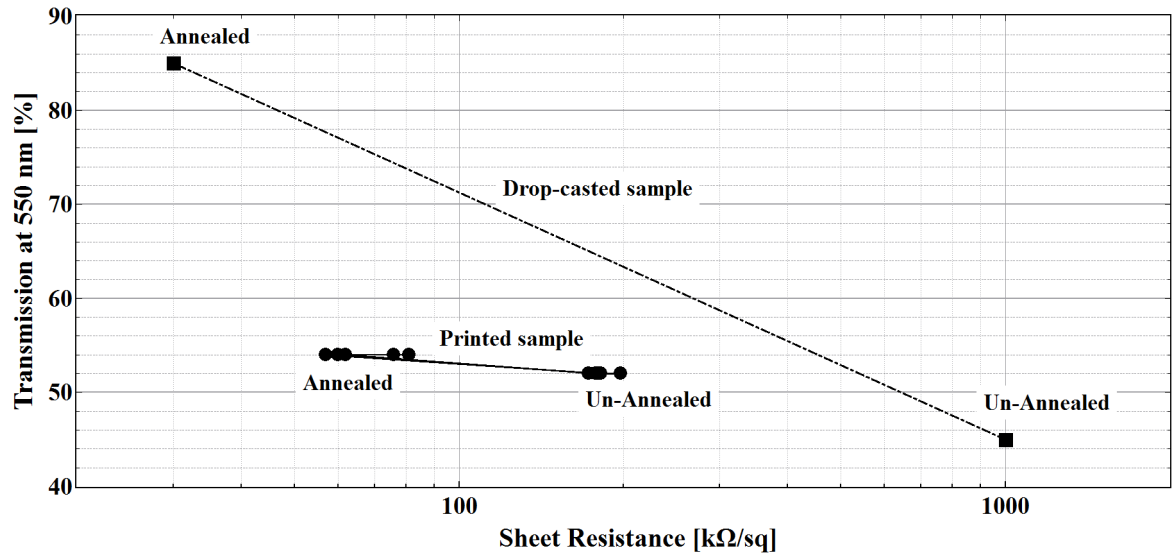

Figure 4. Optical transmission versus sheet resistance before and after laser annealing for drop casted (squares) and inkjet-printed graphene films (dots).

**Table 1:** Comparison of the sheet resistance and transparency of conductive films fabricated from graphene dispersion.

| Method | Sheet Resistance [kΩ/□] | Transparency [%] | Reference |
|---|---|---|---|
| Drop cast + Laser annealing | 30 | 90 | This work |
| Inkjet Printing | 25-175 | 77-88 | [2] |
| Inkjet Printing | 30-100000 | 56-90 | [3] |
| Vacuum filtration | 8-16 | 73 – 85 | [4] |
| Vacuum filtration | 1-5 | 50-75 | [5] |
| Vacuum filtration | 5 | 42 | [6] |
| Spray Coating | 5 | 90 | [7] |

**Acknowledgements**

The authors gratefully acknowledge funding by the European Research Council through the Grant InteGraDe (No. 307311) and the Proof of Concept Grant iPUBLIC (No. 641416), as well as the German Research Foundation (DFG, LE 2440/1-1), the Swedish Research Council through the grant iGRAPHENE and the Göran Gustafsson Foundation through the Young Researcher Prize (No. 1415 B).